\documentclass[twocolumn]{article}

\usepackage{physics}
\usepackage{mathtools}
\usepackage{amssymb} 
\usepackage{xfrac}	
\usepackage{siunitx}	
\usepackage{graphicx}	
\usepackage{subcaption}	
\usepackage[style=phys]{biblatex}	
\usepackage[utf8]{inputenc}	
\usepackage{verbatim}	
\usepackage{abstract}	
\usepackage[hidelinks]{hyperref}	

\graphicspath{{./Pictures/}} 
\newcommand{\Sstates}{$^{2}\text{S}_{1/2}$}	
\newcommand{\mSstates}{{} ^{2}\text{S}_{1/2}}	
\newcommand{\Pstates}{$^{2}\text{P}_{1/2}$}	
\newcommand{\mPstates}{{}^{2}\text{P}_{1/2}}	
\newcommand{\Dstates}{$^{2}\text{D}_{3/2}$}	
\newcommand{\mDstates}{{}^{2}\text{D}_{3/2}}	
\newcommand{\Bstates}{$^{3}\text{D}\qty[\sfrac{3}{2}]_{1/2}$}	
\newcommand{\mBstates}{{}^{3}\text{D}\qty[\sfrac{3}{2}]_{1/2}}	

\newcommand{\Yb}[1][]{$^{#1}$Yb$^{+}$}	
\newcommand{\cool}{$\SI{369}{ \nano\meter}$}
\newcommand{\repump}{$\SI{935}{ \nano\meter}$}

\newcommand{\mTrapRF}{\Omega_{\text{RF}}}	
\newcommand{\Freq}[1]{\omega_{\text{#1}}}	
\newcommand{\Rabi}[1]{\Omega_{\text{#1}}}	
\newcommand{\Detuning}[1]{\Delta_{\text{#1}}}	

\begin{document}
	
	\title{Resolved-sideband Micromotion Sensing in Yb$^{+}$ on the 935 nm Repump Transition}
	\author{Connor J. B. Goham, Joseph W. Britton}
	
	\twocolumn[
	
		\maketitle
		
		\begin{onecolabstract}
			Ions displaced from the potential minimum in a RF Paul trap exhibit excess micromotion. A host of well-established techniques are routinely used to sense (and null) this excess motion in applications ranging from quantum computing to atomic clocks. The rich atomic structure of the heavy ion \Yb\ includes low-lying \Dstates\ states that must be repumped to permit Doppler cooling, typically using a \repump\ laser coupled to the \Bstates\ states. In this manuscript we demonstrate the use of this transition to make resolved-sideband measurements of 3D micromotion in $^{172}$Yb$^{+}$ and $^{171}$Yb$^{+}$ ions. Relative to other sensing techniques our approach has very low technical overhead and is distinctively compatible with surface-electrode ion traps.
		\end{onecolabstract}
	
	]
	
	\section{Introduction}
	Trapped atomic ions are an excellent platform for optical clocks, tests of fundamental physics \cite{countsEvidenceNonlinearIsotope2020}, and as qubits in near-term quantum computing devices \cite{wrightBenchmarking11qubitQuantum2019} \cite{eganFaulttolerantControlErrorcorrected2021}. Of particular interest amongst commonly trapped ionic species are the isotopes of Yb$^{+}$, especially $^{171}$Yb$^{+}$ as it has a spin-$\sfrac{1}{2}$ nucleus with a simple hyperfine qubit for straightforward laser-cooling, spin-state initialization, and readout \cite{olmschenkManipulationDetectionTrapped2007}. The magnetic field insensitive clock states of the HF qubit exhibit long-lived coherence \cite{wangSingleIonQubit2021}. The presence of two narrow quadrupole transitions, as well as one ultra narrow octupole transition, make it a popular approach to building optical atomic clocks \cite{huntemannHighAccuracyOpticalClock2012a}.
	
	Intrinsic to ions trapped in Paul traps is micromotion at the RF trapping frequency $\mTrapRF$. Micromotion along the direction of a probe laser beam causes phase modulation of the laser spectrum, complicating ion-laser interactions including Doppler cooling \cite{berkelandMinimizationIonMicromotion1998}, ion qubit manipulation by Raman-type processes \cite{gaeblerHighFidelityUniversalGate2016}, and probing of narrow clock transitions \cite{kellerPreciseDeterminationMicromotion2015}. When scaling to larger systems these also contribute to spectral crowding \cite{kielpinskiArchitectureLargescaleIontrap2002}, while the motion itself contributes to heating of the ion crystal \cite{chenMeasurementCoulombLogarithm2013} and time-dilation shifts. Furthermore, as it indicates a displacement from the RF null, excess micromotion is associated with Stark shifts. Micromotion sensing and nulling is  therefore a common experimental subroutine across a range of applications.
	
	The most straightforward method for measuring micromotion  is by looking for sidebands on the Doppler cooling transition. In low mass ions such as Be$^{+}$ or Mg$^{+}$, the usual trap RF frequencies, $\sfrac{\mTrapRF}{2\pi}$, are much greater than the D1 linewidth $\sfrac{\Gamma_{\text{D1}}}{2\pi}$, and the micromotion sidebands are well-resolved. For heavier ions such as Yb$^{+}$ or Ba$^{+}$, $\mTrapRF\sim\Gamma_{\text{D1}}$ resulting in poorly resolved sidebands. Furthermore, the probe beam must be tuned over several multiples of $\sfrac{\mTrapRF}{2\pi}$ to unambiguously determine the modulation depth. Given that ion D1 lines are at blue or UV wavelengths, and that the required tuning range approaches or exceeds the bandwidth of available AOMs, this is technically inconvenient in day to day operations. Other techniques include monitoring the time-dependence of ion fluorescence for correlations with the trap RF period \cite{berkelandMinimizationIonMicromotion1998} \cite{nadlingerMicromotionMinimisationSynchronous2021}, or by attempting to drive coherent  operations on the micromotion sidebands of very narrow transitions \cite{kellerPreciseDeterminationMicromotion2015}. In the case of the former, precise time tagging of photon detection events with nanosecond resolution is required. In the case of the latter, narrow lasers are needed. Full determination (and nulling) of micromotion requires measuring in three independent directions. This is complicated by many of the relevant transitions being in the UV, or at the same wavelength as used for detection. With this comes the concern that adding additional probe beams increases the amount of background scatter, adding to detection noise as well as inducing unwanted charging.
	
	In Yb$^{+}$, leakage from the primary laser cooling transition to meta-stable states in \Dstates\ (Figure \ref{fig:LevelStructure}) is repumped through a higher lying manifold of states, forming an effective cycling transition. This transition is both an order of magnitude narrower than the primary cooling transition and smaller than the most commonly used trapping frequencies. As it plays a secondary role in cooling, this transition allows micromotion sidebands to be directly resolved in the cooling fluorescence without the overhead of coherently driving a clock-like transition. Being in the IR, concerns of background scattering hampering the signal to noise or inducing charging are essentially eliminated. This drastically eases alignment constraints for probing in three dimensions, a particular challenge for surface traps. Furthermore, minimal additional hardware is required beyond the standard components necessary for trapping $^{171}$Yb$^{+}$. The few that are required are cheap, off the shelf IR fiber components. This presents a straight forward method of rapid 3D micromotion measurement for day to day calibrations, particularity in cases such as surface traps and integrated optical cavities where nearby surfaces would be susceptible to induced charging from additional UV beams.
	
	\begin{figure}
		\centering
		\begin{subfigure}{\linewidth}
			\centering
			\includegraphics[width=.75\linewidth]{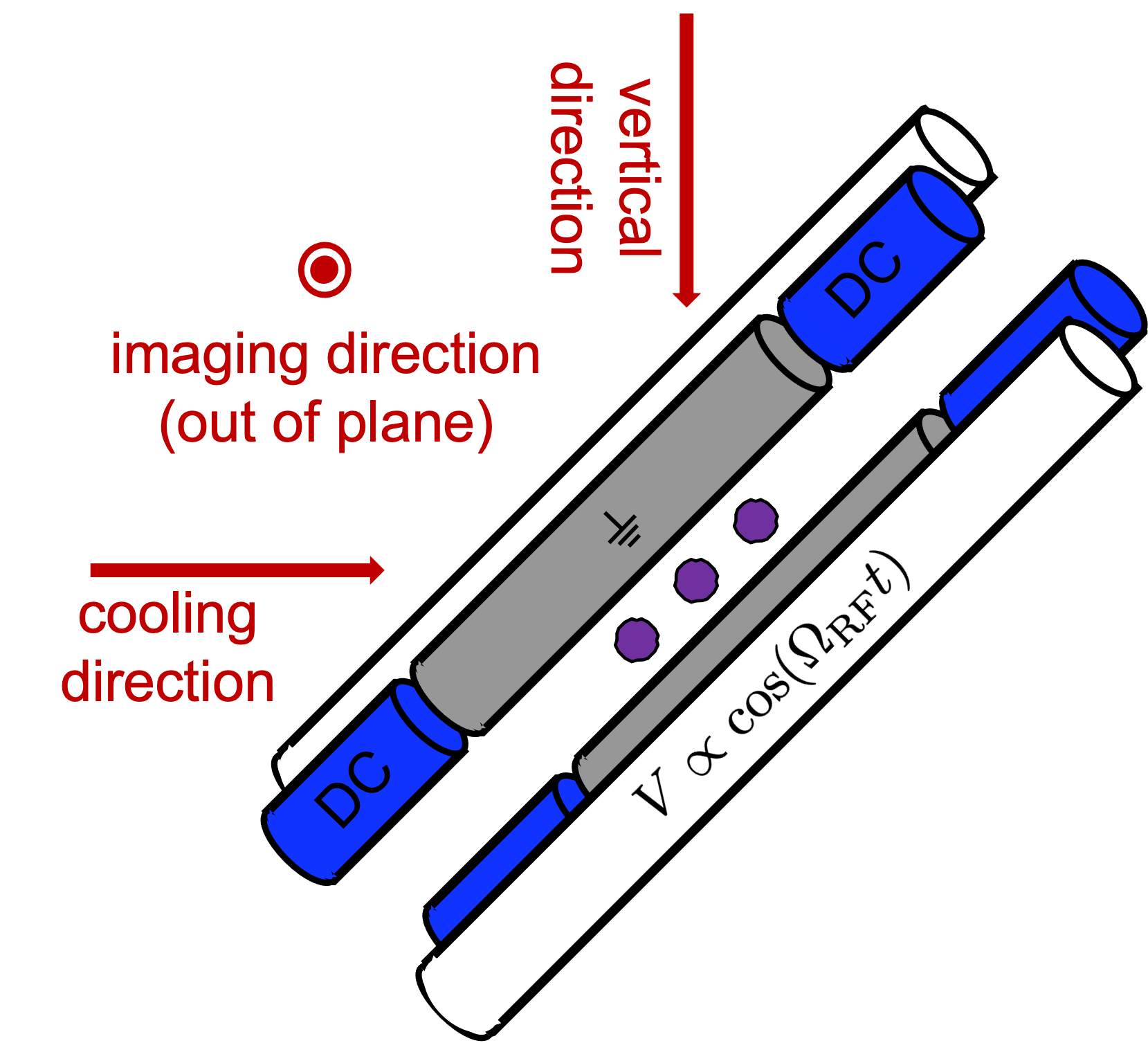}
			\caption{}
			\label{fig:PaulTrap:PerspectiveView}
		\end{subfigure}
		\begin{subfigure}{\linewidth}
			\centering
			\includegraphics[width=.95\linewidth]{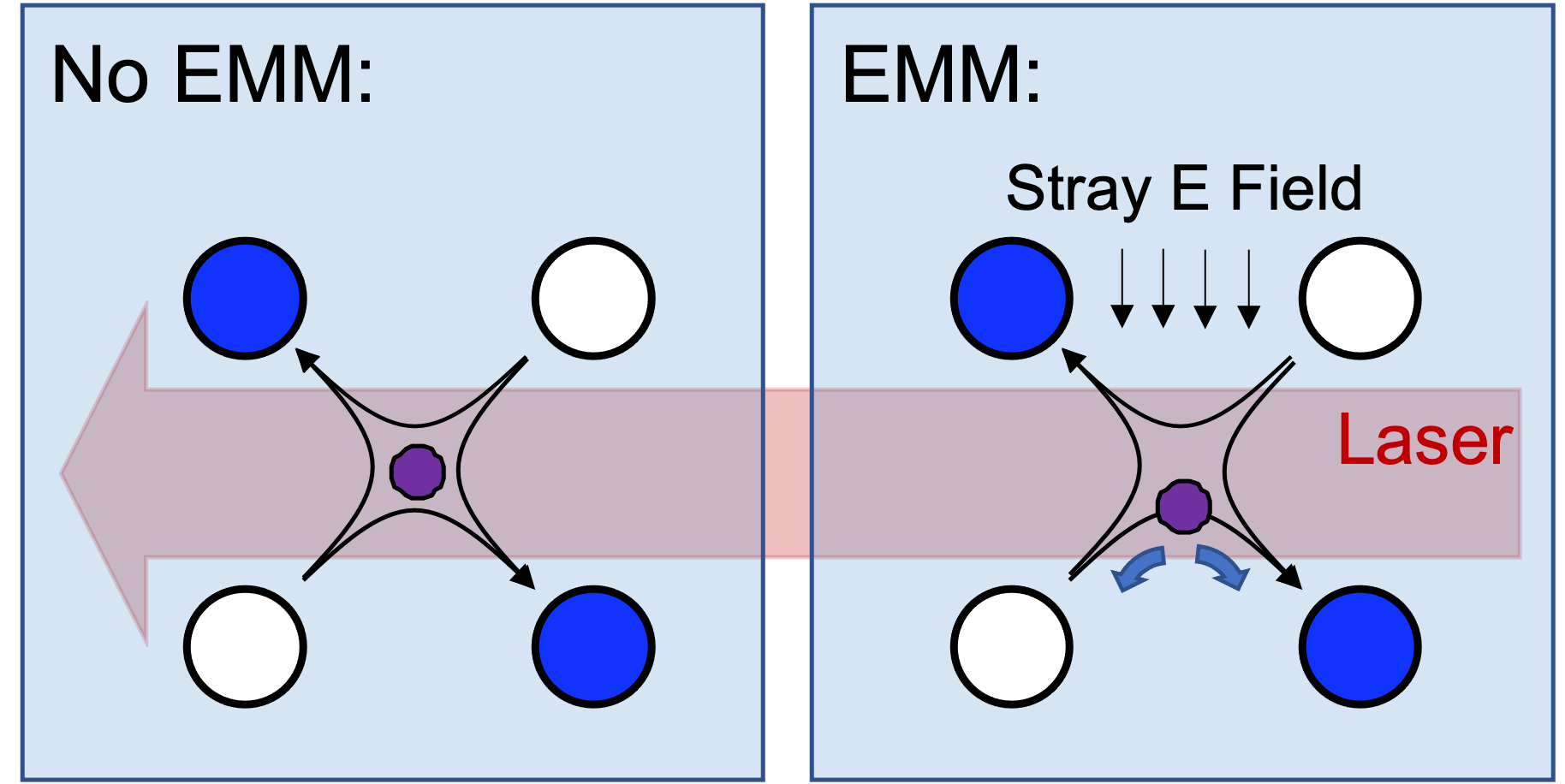}
			\caption{}
			\label{fig:PaulTrap:EMM}
		\end{subfigure}
		\caption{(\ref{fig:PaulTrap:PerspectiveView}) Schematic of a linear RF Paul trap similar to the one used in this experiment. The direction of the probe beams to measure micromotion are shown. (\ref{fig:PaulTrap:EMM}) One cause of excess micromotion (EMM) is the presence of a stray DC field offsetting the equilibrium ion position from the RF null, driving coherent oscillations at the trap RF frequency $\sfrac{\mTrapRF}{2\pi}$. The projection of this motion onto an addressing laser beam results in a phase modulation of the laser as seen by the ion(s).}
		\label{fig:PaulTrap}
	\end{figure}
	
	\section{Methods}
\begin{figure}
		\includegraphics[width=\linewidth]{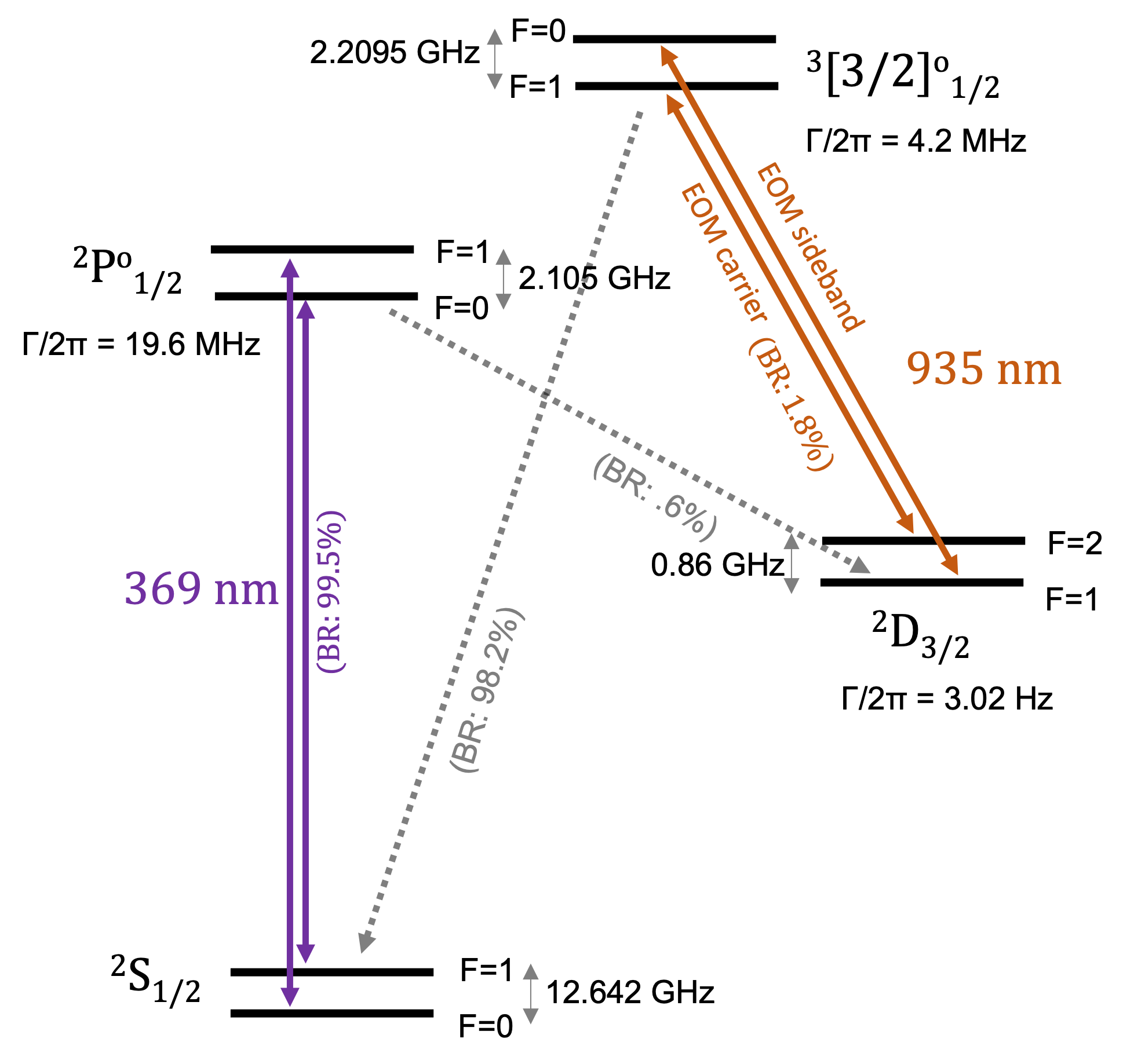}
		\caption{Relevant levels of $^{171}$Yb$^{+}$. The first blue sideband on the \repump\ laser provides primary repumping by cleaning out the $F=1$ hyperfine states of the \Dstates\ manifold. These are the primary states leaked to during cooling. In even isotopes there is no hyperfine structure, and the EOM carrier on the \repump\ is parked far detuned from any transition while a sideband provides repumping. Use of the sideband for primary repumping allows agile control of the laser detuning and strength when probing the lineshape.}
		\label{fig:LevelStructure}
	\end{figure}

	We load ytterbium ions into a blade trap analogous to the diagram in Figure \ref{fig:PaulTrap:PerspectiveView}, with a trap frequency $\mTrapRF=2\pi\times\SI{33.2}{\mega\hertz}$. Copropagating \cool\ light, \repump\ light, as well as light involved in isotope selective photoionization are sent along the ``cooling'' direction in order to load and Doppler cool the ions along all three principal directions. The relevant level structure of a Yb$^{+}$ ion can be seen in Figure \ref{fig:LevelStructure}. Doppler cooling is achieved on the strong $\mSstates\leftrightarrow\mPstates$ transition at $\SI{369}{\nano\meter}$ \cite{olmschenkManipulationDetectionTrapped2007}. The \Pstates\ can also decay to \Dstates\ via a dipole allowed transition at $\SI{2.44}{\micro\meter}$, which happens about once per every 200 scattering events on the primary cooling transition. The ion is returned to the cooling cycle by application of laser light at $\SI{935}{\nano\meter}$ driving the $\mDstates\leftrightarrow\mBstates$ transition, where the \Bstates\ states decay predominantly to the \Sstates\ manifold.
	
	For the $^{171}$Yb$^{+}$ isotope, the cooling cycle stays primarily confined to the $\mSstates\ F=1\leftrightarrow\mPstates\ F=0$ hyperfine levels. The only leakage channel directly allowed by selection rules is to the $F=1$ hyperfine states of the \Dstates\ manifold. This is repumped through the $F=0$ state of the \Bstates\ manifold, closing the cycle. Leakage to other hyperfine states in the \Sstates\  and \Dstates\ manifolds arises from off resonant scatter, namely from the cooling beam scattering off the unintended $\mSstates\ F=1 \leftrightarrow\mPstates\ F=1$ transition detuned by $\SI{2.1}{\giga\hertz}$. Sidebands at approximately $\SI{14.7}{\giga\hertz}$ and $\SI{3.1}{\giga\hertz}$ are added to the \cool\ and \repump\ beams, respectively, restoring population in the cooling cycle. A small magnetic field of 4 gauss is applied to lift the Zeeman degeneracy and prevent population trapping in coherent dark states \cite{berkelandDestabilizationDarkStates2002} \cite{ejtemaeeOptimizationYbFluorescence2010}.
	
	In contrast to the \cool\ transition with a linewidth of $\Gamma_{P}=\SI{19.6}{\mega\hertz}$, the \repump\ transition has a natural linewidth of $\Gamma_{B}=\SI{4.2}{\mega\hertz}$, an order of magnitude lower than both the cooling linewidth and the trap RF. When the repump light is far detuned, the branching ratio is sufficient to rapidly pump the ion into the \Dstates, causing the ion to go dark. In contrast, when an excitation of the $\mDstates\leftrightarrow\mBstates$ does occur, the ion is almost immediately returned to the cooling cycle, yielding fluorescence. As a result the lineshape of the repump transition, with micromotion sidebands spectrally well resolved, can be probed by monitoring cooling fluorescence. This is without driving a narrow clock like transition, and only involves varying the frequency and intensity of the \repump\ light.
	
	In the absence of micromotion and approximating level structure as a four level system, the dependence of the P state population, and hence the detected \cool\ fluorescence, on the cooling and repump laser parameters is given by
	\begin{equation}
		P_{P}=\frac{\sfrac{s_{c}s_{r}}{2}}{s_{c}\qty(\sfrac{\Gamma_{PD}}{\Gamma_{BS}})\qty(1+\delta_{r}^{2}+s_{r})+s_{r}\qty(1+\delta_{c}^{2}+s_{c})}.
		\label{eq:4LevelYbLineshape}
	\end{equation}
	In the above expression, $s_{c}$ and $s_{r}$ are the on resonant saturation parameters of the cooling and repump beams, $\delta_{c}=\sfrac{2\Delta_{c}}{\Gamma_{P}}$ and $\delta_{r}=\sfrac{2\Delta_{r}}{\Gamma_{B}}$ are their detunings normalized to the \Pstates\ and \Bstates\ total scattering rates $\Gamma_{P}$ and $\Gamma_{B}$, and $\Gamma_{ij}$ is the scattering rate from state $i$ to $j$. See the Appendix for a justification of this lineshape. For convenience, ``B'' is used to denote the \Bstates\ ``bracket'' state. In the limit of a weak repumping relative to leakage into the \Dstates\ states, $s_{r}\ll s_{c}\qty(\sfrac{\Gamma_{PD}}{\Gamma_{BS}})$, this can be approximated to first order in $s_{r}$ as
	\begin{equation}
		P_{P}\approx\frac{1}{2}\frac{\Gamma_{BS}}{\Gamma_{PD}}\frac{s_{r}}{1+\delta_{r}^{2}}.
		\label{eq:4LevelYbLineshape:WeakRepump}
	\end{equation}
	
	With the addition of micromotion sidebands on the repump laser, the form of the fluorescence lineshape becomes
	\begin{equation}
		P_{P}\propto\sum_{n=-\infty}^{\infty}\frac{J_{n}^{2}\qty(\beta)}{\qty(\Delta_{r}+n\mTrapRF)^{2}+\qty(\sfrac{\Gamma_{B}}{2})^{2}}.
		\label{eq:2LevelMicromotionLineshape}
	\end{equation}
	$\beta$ can either be extracted from a fit of this lineshape, or if it is sufficiently small and the sidebands well resolved, directly from the ratio of the measured fluorescence on the carrier and first sideband, $\sfrac{\beta^{2}}{4}\approx\sfrac{J_{1}^{2}\qty(\beta)}{J_{0}^{2}\qty(\beta)}$.
	
	To measure the amplitude of micromotion in 3D, we extend the directions we can probe with \repump\ light by adding two additional beam paths, one vertical and one colinear with the imaging objective. A MEMs-mirror-based fiber switch controls which direction the \repump\ is applied along, while a fiber EOM adds sidebands which are rapidly tunable over many GHz. A fiber AOM is used to dynamically control the amplitude; set to full during standard cooling, and attenuated to be below saturation when probing micromotion.
	
	The technique was first validated using $^{172}$Yb$^{+}$ as its D1 transition has no coherent dark states. This allows the ion to be trapped at zero magnetic field, in turn preventing  any Zeeman splitting on the $\mDstates\leftrightarrow\mBstates$ transition, keeping the micromotion sidebands maximally resolved. In order to provide agile control of the probe light, the \repump\ laser was tuned so that the fiber EOM carrier was detuned $\SI{10}{\giga\hertz}$ red of the desired transition. The first blue sideband was used to provide repumping, allowing fast tuning of the repump color over a large detuning range.
	
	For micromotion detection directly with $^{171}$Yb$^{+}$, the carrier of the EOM was set on resonance with the $\mDstates\ F=2\leftrightarrow\mBstates\ F=1$ transition, preventing population buildup from off resonant scattering into this state. The first blue sideband of the EOM was set to span the hyperfine splitting, driving the $\mDstates\ F=1\leftrightarrow\mBstates\ F=0$ transition. The magnetic field was increased to approximately 4 G to prevent coherent dark states and resulting in Zeeman broadening of the transitions, however, micromotion sidebands remain resolvable given the $\SI{33.2}{\mega\hertz}$ trap frequency.
	
	Using the measured micromotion as feedback, the trapping potentials were then modified to reduce EMM in simultaneously in all three directions. This was facilitated by independent control of the DC voltages applied to all four of the end cap electrodes on the two DC blades (analogous to the four blue electrodes labeled ``DC'' in diagram \ref{fig:PaulTrap:PerspectiveView}) as well as the to the center DC electrodes (analogous to the gray grounded electrodes in \ref{fig:PaulTrap:PerspectiveView}). In addition, a DC bias could be added to each of the RF blades. Control of all of the trapping potentials, lasers, and measurements was performed using ARTIQ \cite{sebastienbourdeauducqMlabsArtiq2021}.
	
	\section{Results}
\begin{figure}
		\centering
		\begin{subfigure}{\linewidth}
			\centering
			\includegraphics[width=.95\linewidth]{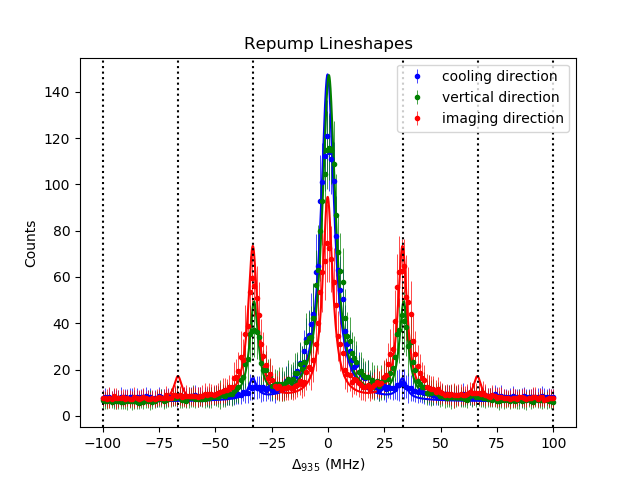}
			\caption{Excess micromotion present}
			\label{fig:172:Measured3DLineshapes:EcxessMicromotion}
		\end{subfigure}
		\begin{subfigure}{\linewidth}
			\centering
			\includegraphics[width=.95\linewidth]{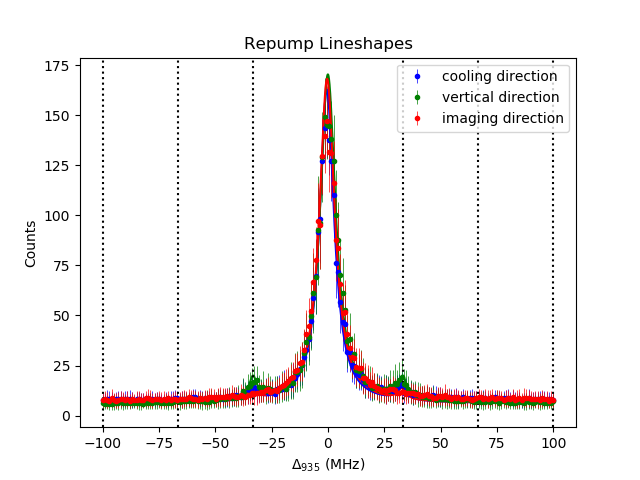}
			\caption{Reduced excess micromotion}
			\label{fig:172:Measured3DLineshapes:ReducedMicromotion}
		\end{subfigure}
		\caption{Measured lineshape of the \repump\ repump transition in $^{172}$Yb$^{+}$ when probed along three orthogonal directions (one copropagating with the cooling beam, one oriented vertically, and one co-linear with the imaging axis). (\ref{fig:172:Measured3DLineshapes:EcxessMicromotion}) After minimization of micromotion along the cooling direction using the standard photon correlation method, the 3D measurement reveals a large amount of EMM present in the plane orthogonal to the cooling beam. (\ref{fig:172:Measured3DLineshapes:ReducedMicromotion}) Micromotion is reduced in 3D by adjusting the compensation voltages while probing the lineshape along all three directions. Dashed vertical lines guide the eye as to where the micromotion sidebands are expected to occur. Residual micromotion present may be the result of trap imperfections preventing full nulling.}
		\label{fig:172:Measured3DLineshapes}
	\end{figure}

	The effective lineshape of the $\mDstates\leftrightarrow\mBstates$ transition of $^{172}$Yb$^{+}$ along three independent directions, observed as the efficacy of restoring fluorescence at \cool\ for various \repump\ detunings, can be seen in Figure \ref{fig:172:Measured3DLineshapes}. The necessity of a full 3D measurement when minimizing micromotion is made clear by Figure \ref{fig:172:Measured3DLineshapes:EcxessMicromotion}. The trapping potentials used when this lineshape was taken were arrived at by first minimizing micromotion along the direction of the cooling beam with the standard photon correlation method at \cool. Although successful in that direction, large amounts of micromotion remained present in the orthogonal plane; making clear that reduced micromotion along one direction is not a sufficient proxy for the for the other two.
	
\begin{figure}
		\centering
		\begin{subfigure}{\linewidth}
			\centering
			\includegraphics[width=.95\linewidth]{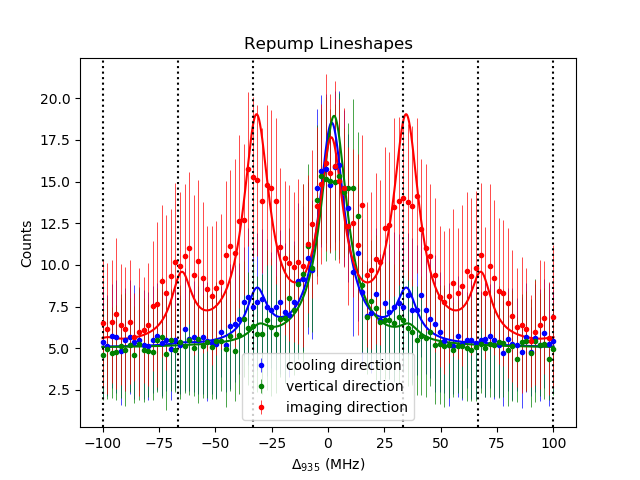}
			\caption{Excess micromotion present}
			\label{fig:171:Measured3DLineshapes:EcxessMicromotion}
		\end{subfigure}
		\begin{subfigure}{\linewidth}
			\centering
			\includegraphics[width=.95\linewidth]{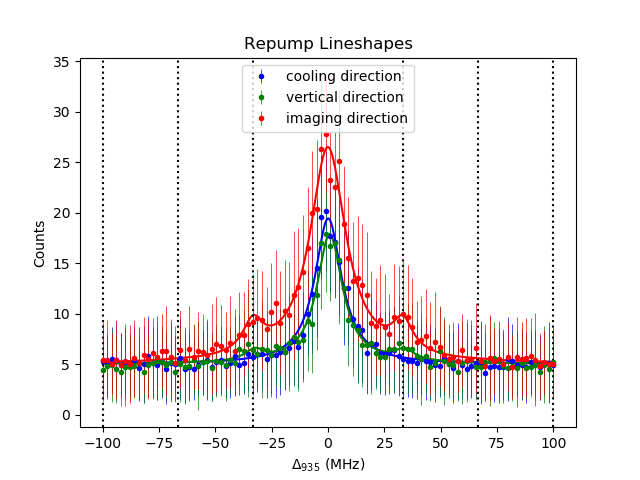}
			\caption{Reduced excess micromotion}
			\label{fig:171:Measured3DLineshapes:ReducedMicromotion}
		\end{subfigure}
		\caption{Measured lineshape of the \repump\ repump transition in $^{171}$Yb$^{+}$ when probed along three orthogonal directions (same directions as for $^{172}$Yb$^{+}$ in Figure \ref{fig:172:Measured3DLineshapes}). The addition of hyperfine structure in $^{171}$Yb$^{+}$ results in coherent dark states in the cooling cycle, requiring a small B field to destabilize via lifting the Zeeman degeneracy. These result in reduced fluorescence and broadened lines when compared to $^{172}$Yb$^{+}$. Dashed vertical lines guide the eye as to where the micromotion sidebands are expected to occur.}
		\label{fig:171:Measured3DLineshapes}
	\end{figure}

	Similarly, the 3D lineshape can be taken for $^{171}$Yb$^{+}$, as shown in Figure \ref{fig:171:Measured3DLineshapes}. Fluorescence from this isotope is reduced relative to $^{172}$Yb$^{+}$ due the presence of coherent dark states on  the $\mSstates\ F=1\leftrightarrow\mPstates\ F=0$ primary cooling transition \cite{berkelandDestabilizationDarkStates2002}. Being able to measure micromotion directly with $^{171}$Yb$^{+}$, however, is desired since this is often the isotope of interest.
	
	Fitting lineshapes of the form of Eq. (\ref{eq:2LevelMicromotionLineshape}), the modulation depths along all three directions were extracted. The results for the two isotopes can be seen in Tables \ref{table:172:ModulationDepths} and \ref{table:171:ModulationDepths}, respectively, with the minimum achieved  modulation depth corresponding to a sensitivity of \SI{1150}{\volt\per\meter}. Differences between the modulation depths measured for the two isotopes can be explained by charging in between measurements, as the measurements  were taken at different times and charging had been observed to occur, particularity after alignment of the various UV beams.
	
	\begin{table}
		\centering
		\caption{Measured modulation index, $\beta$, of the \repump\ beam along each of the directions for $^{172}$Yb$^{+}$.}
		\begin{tabular}{r c c}
			\hline\hline
			direction & With EMM & Reduced EMM \\
			\hline
			cooling & $0.30 \pm 0.03$ & $0.25 \pm 0.02$\\
			vertical & $0.68 \pm 0.02$ & $0.27 \pm 0.02$\\
			imaging & $1.27 \pm 0.02$ & $0.19 \pm 0.03$\\
			\hline\hline
		\end{tabular}
		\label{table:172:ModulationDepths}
	\end{table}
	
	\begin{table}
		\centering
		\caption{Measured modulation index, $\beta$, of the \repump\ beam along each of the directions for $^{171}$Yb$^{+}$.}
		\begin{tabular}{r c c}
			\hline\hline
			direction & With EMM & Reduced EMM \\
			\hline
			cooling & $0.70 \pm 0.03$ & $0.025 \pm 0.02$\\
			vertical & $0.36 \pm 0.07$ & $0.43 \pm 0.04$\\
			imaging & $1.53 \pm 0.04$ & $0.42 \pm 0.02$\\
			\hline\hline
		\end{tabular}
		\label{table:171:ModulationDepths}
	\end{table}
	
	\section{Conclusion}
	Probing the \repump\ repump lineshape of trapped Yb$^{+}$ ions allows rapid detection of micromotion in 3D, including in directions otherwise inaccessible to UV lasers used in standard measurement techniques. Issues of induced charging are reduced, precise photon timing is not required, and measurement does not depend on coherent control. Well developed NIR fiber technology is available, providing agile control of the probe beams for rapid measurements. Furthermore, because the detected signal is the ratio of fluorescence observed on the sidebands relative to the carrier, this method readily extends to large crystals of ions that may have differing absolute count rates due to spatial gradients in the beam intensities. If site resolved imaging and readout is available, a full map of the 3D micromotion can be simultaneously mapped across an entire crystal.
	
	The convenience of this method allows integration as a rapid calibration subroutine that can be incorporated in day to day operation of Yb$^{+}$ systems. This will be particularly advantageous in future experiments where the ion(s) are located near dielectric surfaces with time-varying stray charges, such as mirrors of small volume optical cavities \cite{steinerPhotonEmissionAbsorption2014}. It may also serve helpful in efforts to extend ion crystals into two dimensions, where excess micromotion along some directions becomes unavoidable \cite{donofrioRadialTwoDimensionalIon2021} \cite{yoshimuraCreationTwodimensionalCoulomb2015}.
	
	\appendix
	
	\section{Repump Lineshape}
	The expected lineshape (\ref{eq:4LevelYbLineshape}) is derived by approximating Yb$^{+}$ as a four level system and solving the Lindblad Master equation. Considering the undriven eigenstates as a basis, $\mathcal{B}=\qty(\ket{\text{P}},\ket{\text{S}},\ket{\text{B}},\ket{\text{D}})$, the undriven Hamiltonian is
	\begin{equation}
		\hat{H}_{a}=\sum_{i\in\mathcal{B}}\hbar\omega_{i}\ketbra{i}.
		\label{eq:Appendix:4LevelUndrivenLabHamiltonian}
	\end{equation}
	For  ``cooling'' and ``repump'' lasers driving transitions between $\text{S}\leftrightarrow\text{P}$ and $\text{D}\leftrightarrow\text{B}$, respectively, the Hamiltonian for the drive fields under the rotating wave approximation (RWA) is
	\begin{equation}
		\hat{H}_{la}=\frac{\hbar}{2}\mqty[
		0 & \Rabi{c}e^{-i\Freq{c}t} & 0 & 0\\
		\Rabi{c}e^{i\Freq{c}t} & 0 & 0 & 0\\
		0 & 0 & 0 & \Rabi{r}e^{-i\Freq{r}t}\\
		0 & 0 & \Rabi{r}e^{i\Freq{r}t} & 0\\
		]
		\label{eq:Appendix:4LevelDriveHamiltonian}
	\end{equation}
	Moving to a rotating frame by making the transformation $\hat{\mathcal{R}}=\exp[i\qty(\hat{G}+\frac{1}{\hbar}\hat{H}_{a})t]$ where
	\begin{equation}
		\hat{G}=\frac{1}{2}\mqty[\dmat[0]{\Detuning{c},-\Detuning{c},\Detuning{r},-\Detuning{r}}],
		\label{eq:Appendix:DriveFrameRotationGenerator}
	\end{equation}
	the full Hamiltonian under the rotating wave approximation becomes time independent:
	\begin{equation}
		\hat{H}=\hat{H}_{a}+\hat{H}_{la}\longrightarrow\frac{\hbar}{2}
		\mqty[
			-\Delta_{c} & \Omega_{c} & 0 & 0\\
			\Omega_{c} & \Delta_{c} & 0 & 0\\
			0 & 0 & -\Delta_{r} & \Omega_{r}\\
			0 & 0 & \Omega_{r} & \Delta_{r}\\
		].
		\label{eq:Appendix:4LevelDriveFrameHamiltonian}
	\end{equation}
	The states $\ket{P}$ and $\ket{B}$ are allowed to decay to both $\ket{S}$ and $\ket{D}$, as encoded in the Lindbald superoperator
	\begin{equation}
		\mathcal{D}\qty(\rho)=\sum_{\substack{e\in\qty{P,B}\\ g\in\qty{S,D}}}\Gamma_{eg}\qty(\hat{\sigma}_{ge}\rho\hat{\sigma}_{ge}^{\dagger}-\frac{1}{2}\acomm{\hat{\sigma}_{ge}^{\dagger}\hat{\sigma}_{ge}}{\rho})
		\label{eq:Appendix:DecaySuperOperator}
	\end{equation}
	where $\hat{\sigma}_{ge}=\ketbra{g}{e}$ and $\acomm{\underline{\hphantom{\sigma}}}{\underline{\hphantom{\sigma}}}$ is the anti-commutator. Quadrupole decays of D to S have been ignored for simplicity. This is a safe approximation given that the decay rate of the \Dstates\ is $\Gamma_{DS}\approx\SI{20}{\hertz}$, which is orders of magnitude slower that the other rates involved (see Figure \ref{fig:LevelStructure}). Finding the steady state solution $\tilde{\rho}$ of of the Lindblad Master equation
	\begin{equation}
		\dot{\rho}=\frac{1}{i\hbar}\comm{\hat{H}}{\rho}+\mathcal{D}\qty(\rho),
		\label{eq:MasterEquation}
	\end{equation}
	by setting $\dot{\tilde{\rho}}=0$ and solving the resulting system of equations with the normalization condition $\tr(\tilde{\rho})=1$, the observed fluorescence is proportional to $P_{P}=\tilde{\rho}_{PP}$. The dependence of $P_{P}$ on the \cool\ and \repump\ laser parameters is given by Eq. (\ref{eq:4LevelYbLineshape}).
	
	\printbibliography
	
\end{document}